\documentclass[a4paper,11pt]{article}
\usepackage[left=2cm,right=2cm,top=2cm,bottom=2cm]{geometry}

\usepackage{amsmath,lscape,amsfonts,amssymb,amsthm,verbatim,color,lscape}
\usepackage[figuresright]{rotating}
\usepackage{natbib}
\usepackage{url}
\usepackage{hyperref}
\hypersetup{colorlinks,
citecolor=black,
filecolor=black,
linkcolor=black,
urlcolor=black,
pdftex}

\usepackage{booktabs}

\usepackage{colortbl}
\definecolor{gray}{rgb}{0.8,0.8,0.8}

\usepackage{times}
\usepackage{blindtext}
\usepackage[font={footnotesize,it}]{caption}

\usepackage{sectsty}  
\subsectionfont{\normalsize\bf}
\sectionfont{\large\bf}

\def \red#1{\textcolor{red}{#1}}

\def \red#1{\textcolor{red}{#1}}

\usepackage{titlesec}
\titlespacing\subsubsection{0pt}{2pt plus 4pt minus 2pt}{2pt plus 2pt minus 2pt}

\begin{document}

\def\p{\partial}
\def\oo{\infty}
\def\rt#1{\sqrt{#1}\,}

\def\Cbar{{\overline C}}
\def\C{\mathbf{C}}
\def\E{{\rm E}\,}
\def\I{\mathbf{I}}
\def\pp{\mathbf{p}}
\def\R{\mathbf{R}}
\def\y{\mathbf{y}}
\def\Y{\mathbf{Y}}
\def\z{\mathbf{z}}
\def\x{\mathbf{x}}
\def\o{\omega}
\def\s{\sigma}

\def\V{\mathbf{V}}
\def\I{\mathbf{I}}
\def\bfv{\mathbf{v}}
\def\X{\mathbf{X}}
\def\D{\mathbf{D}}

\def\a{\boldsymbol{\alpha}}
\def\g{\gamma}
\def\b{\beta}

\def\de{\delta}
\def\debf{\boldsymbol{\delta}}
\def\e{\epsilon}
\def\th{\theta}
\def\thbf{\boldsymbol{\theta}}
\def\pibf{\boldsymbol{\pi}}
\def\Xibf{\boldsymbol{\Xi}}
\def\Sbf{\boldsymbol{\Sigma}}

\def \red#1{\textcolor{red}{#1}}
\def \blue#1{\textcolor{blue}{#1}}
\def \magenta#1{\textcolor{magenta}{#1}}
\def \green#1{\textcolor{green}{#1}}
\def\bbf{\boldsymbol{\beta}}
\def\bmu{\boldsymbol{\mu}}

\long\def\symbolfootnote[#1]#2{\begingroup
\def\thefootnote{\fnsymbol{footnote}}\footnote[#1]{#2}\endgroup}
\newcommand{\strike}{\color{red}\sout}

\newcommand{\widesim}[2][1.5]{
  \mathrel{\overset{#2}{\scalebox{#1}[1]{$\sim$}}}
}

\title{On composite likelihood in bivariate meta-analysis of diagnostic test accuracy  studies}

\author{
Aristidis K. Nikoloulopoulos\footnote{{\small\texttt{A.Nikoloulopoulos@uea.ac.uk}}, School of Computing Sciences, University of East Anglia,
Norwich NR4 7TJ, UK}
}
\date{}

\maketitle

\begin{abstract}
\baselineskip=12pt

\noindent The composite likelihood (CL) is amongst the computational methods used for estimation of the generalized linear mixed model (GLMM) in the context of bivariate meta-analysis of diagnostic test accuracy studies. Its advantage is that the likelihood can  be derived conveniently under the assumption of independence between the random effects, but there has not been a clear analysis of the  merit or necessity  of this method. 
For synthesis of diagnostic test accuracy studies, a copula mixed model  has been proposed  in the biostatistics literature.
This general model includes  the GLMM 
 as a special case and  can also  allow for flexible dependence modelling, different from assuming simple linear correlation structures, normality and tail independence in the joint tails. A maximum likelihood  (ML) method, which is based  on evaluating the bi-dimensional integrals of the likelihood with  quadrature methods has been proposed, and in fact   it eases any  computational difficulty that might be caused by the double integral in the likelihood function. 
Both methods are thoroughly examined with extensive simulations and illustrated with data of  a published meta-analysis. It is shown that the ML method has no non-convergence issues or computational difficulties and at the same time  allows estimation of the dependence between study-specific sensitivity and specificity  and thus prediction via summary receiver operating curves.
\\\\
\noindent {\it Keywords:} {Copula mixed model; diagnostic odds ratio; generalized linear mixed model; sensitivity/specificity; SROC.}
\end{abstract}

\baselineskip=13pt

\section{Introduction}

Synthesis of diagnostic test accuracy studies is  the most common medical application of multivariate meta-analysis; we refer the interested reader to  the 
surveys by  \cite{JacksonRileyWhite2011,MavridisSalanti13,Ma-etal-2013}. 
These data have two important properties. The first is that the estimated sensitivities and specificities are typically negatively associated across studies,  because studies that adopt less stringent criterion for declaring a test positive invoke higher sensitivities and lower specificities \citep{JacksonRileyWhite2011}. The second important property of the data is the substantial between-study heterogeneity in sensitivities and specificities \citep{Chu-etal-2012}.

\cite{Nikoloulopoulos2015b}, to deal with the aforementioned  properties,
proposed a copula mixed model for bivariate meta-analysis of diagnostic test accuracy studies and made the argument for moving to copula random effects models. 
This general model includes  the generalized linear mixed model   \citep{Chu&Cole2006,Arends-etal-2008} 
 as a special case and  can also operate on the original scale of sensitivity and specificity.

\cite{Chen-etal-smmr-2014,Chen-etal-sim-2015b} 
proposed a composite likelihood (CL) method  for estimation of  the the generalized linear mixed model (hereafter GLMM) and the Sarmanov beta-binomial model \citep{Chu-etal-2012}. Note in passing that both models are  special cases of a copula mixed model \citep{Nikoloulopoulos2015b}.
The composite likelihood can  be derived conveniently under the assumption of independence between the random effects.  
The CL method has been recommended by  \cite{Chen-etal-smmr-2014,Chen-etal-sim-2015b}  to overcome practical `issues' in the joint likelihood inference such as  computational difficulty caused by a double integral in the joint likelihood function, and restriction to bivariate normality.

However, 
\begin{itemize}
\item[(a)]
GLMM can only be unstable if there are too many parameters in the covariance
matrix of the random effects or too many random effects for a small
sample, which is not the case in this application domain. Furthermore,   \cite{Chen-etal-smmr-2014,Chen-etal-sim-2015b} 
restrict themselves  to SAS PROC NLMIXED which is a general routine for random effect models and thus  gives limited capacity. The CL method is well established as a surrogate alternative of maximum likelihood when the joint likelihood is too difficult to compute \citep{Varin-etal2011}, which is apparently not the case in the synthesis of diagnostic test accuracy studies. 
The  general model in \cite{Nikoloulopoulos2015b} includes the GLMM as a special case and its  
numerical evaluation  
has been  
implemented in the package {\tt CopulaREMADA} \citep{Nikoloulopoulos-2015} within the open source statistical environment {\tt R} \citep{CRAN}.
\item[(b)] The random effects distribution of a copula mixed model can be expressed via other copulas (other than the bivariate normal) that allow for flexible dependence modelling, different from assuming simple linear correlation structures, normality and tail independence. 
\end{itemize}

The contribution of this paper is  to examine the  merit of the CL method in the context of diagnostic test accuracy studies and compare it to the ML method in \cite{Nikoloulopoulos2015b}.
The remainder of the paper proceeds as follows. Section \ref{sec-model} summarizes the copula mixed model for diagnostic test accuracy studies. 
 Section \ref{sec-est} discusses  both maximum and composite likelihood for estimation of the model parameters.  Section \ref{sec-sim} contains  small-sample  efficiency calculations
to  compare the two methods.
Section \ref{sec-app} presents applications  of the likelihood estimation methods  to several data frames with diagnostic studies. We conclude with some discussion in Section \ref{sec-disc}.

\section{\label{sec-model}The copula mixed model}
We first introduce the notation used in this paper. The focus is on two-level (within-study and between-studies) cluster data. The data are are $(y_{ij}, n_{ij}),\, i = 1, . . . ,N,\, j=1,2$, where $j$ is an index for the within study measurements and $i$ is an
index for the individual studies.
The data, for study $i$, can be summarized in a $2\times 2$ table with the number of true positives
($y_{i1}$), true negatives ($y_{i2}$), false negatives ($n_{i1}-y_{i1}$), and false positives ($n_{i2}-y_{i2}$).

The within-study model assumes that the number of true positives $Y_{i1}$ and true negatives $Y_{i2}$ are conditionally independent and binomially distributed given $\X=\x$, where $\X=(X_1,X_2)$ denotes the  bivariate latent  pair of (transformed) sensitivity and specificity.  That is
\begin{eqnarray}\label{withinBinom}
Y_{i1}|X_{1}=x_1&\sim& \mbox{Binomial}\Bigl(n_{i1},l^{-1}(x_1)\Bigr);\nonumber\\
Y_{i2}|X_{2}=x_2&\sim& \mbox{Binomial}\Bigl(n_{i2},l^{-1}(x_2)\Bigr),
\end{eqnarray}
where $l(\cdot)$ is a link function.

The stochastic representation of the between studies model takes the form
\begin{equation}\label{copula-between}
\Bigl(F\bigl(X_1;l(\pi_1),\de_1\bigr),F\bigl(X_2;l(\pi_2),\de_2\bigr)\Bigr)\sim C(\cdot;\th),
\end{equation}
where $C(\cdot;\th)$ is a parametric family of copulas with dependence parameter $\th$ and $F(\cdot;l(\pi),\de)$ is the cdf of the univariate distribution of the random effect. The copula parameter $\th$ is a parameter of the random effects model and it is separated from the univariate parameters, the univariate parameters $\pi_1$ and $\pi_2$ are the meta-analytic parameters for the sensitivity and specificity, and  $\de_1$ and $\de_2$  express the between-study variabilities.
The models in (\ref{withinBinom}) and (\ref{copula-between}) together specify a copula mixed  model \citep{Nikoloulopoulos2015b} with joint likelihood

\begin{equation}\label{mixed-cop-likelihood}
L(\pi_1,\pi_2,\de_1,\de_2,\th)=\prod_{i=1}^N\int_{0}^{1}\int_{0}^{1}
\prod_{j=1}^2g\Bigl(y_{ij};n_{ij},l^{-1}\bigl(F^{-1}(u_j;l(\pi_j),\de_j)\bigr)\Bigr)c(u_1,u_2;\th)du_1du_2,
\end{equation}
where $c(u_1,u_2;\th)=\p^2 C(u_1,u_2;\th)/\p u_1\p u_2$ is the copula  density and $g\bigl(y;n,\pi\bigr)=\binom{n}{y}\pi^y(1-\pi)^{n-y},\quad y=0,1,\ldots,n,\quad 0<\pi<1,$
 is the binomial probability mass function (pmf).
The choices of the  $F\bigl(\cdot;l(\pi),\de\bigr)$ and  $l$ are given in Table \ref{choices}.

\setlength{\tabcolsep}{33pt}
\begin{table}[!h]
\begin{center}
\caption{\label{choices}The choices of the  $F\bigl(\cdot;l(\pi),\de\bigr)$ and  $l$ in the copula mixed model.}
\begin{tabular}{cccc}
\hline $F\bigl(\cdot;l(\pi),\de\bigr)$ & $l$ & $\pi$ & $\de$\\\hline
$N(\mu,\s)$ & logit, probit, cloglog & $l^{-1}(\mu)$&$\s$\\
Beta$(\pi,\gamma)$ & identity & $\pi$ & $\gamma$\\
\hline
\end{tabular}

\end{center}
\end{table}

\section{\label{sec-est}Estimation methods}

\subsection{Maximum likelihood method}
Estimation of the model parameters $(\pi_1,\pi_2,\de_1,\de_2,\th)$    can be approached by the standard maximum likelihood (ML) method, by maximizing the logarithm of the joint likelihood in (\ref{mixed-cop-likelihood}).
For mixed models of the form with joint likelihood as in (\ref{mixed-cop-likelihood}), numerical evaluation of the joint pmf is easily done with the following steps \citep{Nikoloulopoulos2015b}:

\begin{enumerate}
\item Calculate Gauss-Legendre  quadrature points $\{u_q: q=1,\ldots,n_q\}$ 
and weights $\{w_q: q=1,\ldots,n_q\}$ in terms of standard uniform distribution \citep{Stroud&Secrest1966}. Our  comparisons with more
quadrature points show that $n_q=15$ is adequate with good precision to at least at four decimal places \cite[Appendix]{Nikoloulopoulos2015b}.
\item Convert from independent uniform random variables $\{u_{q_1}: q_1=1,\ldots,n_q\}$ and $\{u_{q_2}: q_2=1,\ldots,n_q\}$ to dependent uniform random variables $\{u_{q_1}: q_1=1,\ldots,n_q\}$ and $\{C^{-1}(u_{q_2}|u_{q_1};\th): q_1=q_2=1,\ldots,n_q\}$ that have distribution $C(\cdot;\th)$.
The inverse of the conditional distribution $C(v|u;\th)=\partial C(u,v;\th)/\partial u$ corresponding to the copula $C(\cdot;\th)$ is used  to achieve this.
\item Numerically evaluate the joint pmf
$$\int_{0}^{1}\int_{0}^{1}
\prod_{j=1}^2g\Bigl(y_{ij};n_{ij},l^{-1}\bigl(F^{-1}(u_j;l(\pi_j),\de_j)\bigr)\Bigr)c(u_1,u_2;\th)du_1du_2$$
in a double sum:
\end{enumerate}
$$\sum_{q_1=1}^{n_q}\sum_{q_2=1}^{n_q}w_{q_1}w_{q_2}
g\Bigl(y_{1};n_1,l^{-1}\bigr(F^{-1}(u_{q_1};l(\pi_1),\g_1)\bigr)\Bigr)g\Bigl(y_{2};n_2,l^{-1}\bigl(F^{-1}(C^{-1}(u_{q_2}|u_{q_1};\th);l(\pi_2),\g_2)\bigr)\Bigr).$$

The inverse conditional copula cdfs  $C^{-1}(v|u;\th)$ are given in Table \ref{2fam} for   the  sufficient list of parametric families of copulas  for meta-analysis of diagnostic test accuracy studies in  \cite{Nikoloulopoulos2015b,Nikoloulopoulos2015c}. Since  the copula parameter $\th$ of each   family has different range, in the sequel we re-parametrize them via their Kendall’s $\tau$; that is comparable across families.

\setlength{\tabcolsep}{3pt}
\begin{table}[!h]
\caption{\label{2fam}Parametric families of bivariate copulas and their Kendall's $\tau$ as a strictly increasing function of the copula parameter $\theta$.}
\begin{small}
\centering
\begin{tabular}{ccc}
\hline
Copula & $C^{-1}(v|u;\th)$& 
$\tau$\\\hline
BVN & $\Phi\Bigl(\sqrt{1-\th^2}\Phi^{-1}(v)+\th\Phi^{-1}(u)\Bigr)$ 
&$\frac{2}{\pi}\arcsin(\th)\quad ,\quad -1\leq\th\leq1$\\
Frank &$
-\frac{1}{\theta}\log\left[1-\frac{1-e^{-\th}}{(v^{-1}-1)e^{-\th u}+1}\right]
$
&$\begin{array}{ccc}
1-4\theta^{-1}-4\theta^{-2}\int_\theta^0\frac{t}{e^t-1}dt &,& \th<0\\
1-4\theta^{-1}+4\theta^{-2}\int^\theta_0\frac{t}{e^t-1}dt &,& \th>0\\
\end{array}$\\
Clayton  &$\Bigl\{(v^{-\theta/(1+\theta)}-1)u^{-\th}+1\Bigr\}^{-1/\theta}$
 &$\th/(\th+2)\quad ,\quad \th>0$\\
Clayton by 90 &$\Bigl\{(v^{-\theta/(1+\theta)}-1)(1-u)^{-\th}+1\Bigr\}^{-1/\theta}$&
$-\th/(\th+2)\quad ,\quad \th>0$\\
Clayton by 180 &$1-\Bigl[\bigl\{(1-v)^{-\theta/(1+\theta)}-1\bigr\}(1-u)^{-\th}+1\Bigr]^{-1/\theta}$
&$\th/(\th+2)\quad ,\quad \th>0$\\
Clayton by 270 &$1-
\Bigl[\bigl\{(1-v)^{-\theta/(1+\theta)}-1\bigr\}u^{-\th}+1\Bigr]^{-1/\theta}$
&$-\th/(\th+2)\quad ,\quad \th>0$\\
\hline
\end{tabular}
\end{small}
\end{table}

\subsection{Composite likelihood method}
The composite likelihood method assumes  independence between the random effects.  Hence, it is identical  for  any copula mixed model, since all the parametric families of copulas in Table   \ref{2fam} contain the independence copula as a special case. 
This subsection summarizes the composite likelihood estimating equations and  the asymptotic covariance matrix for the estimator that solves them in the context of diagnostic test accuracy studies.
\subsubsection{Composite likelihood estimator}
\cite{Chen-etal-smmr-2014} and \cite{Chen-etal-sim-2015b} proposed the composite likelihood method for estimation of the copula mixed model  with normal and beta margins, respectively. 
Composite likelihood is  a surrogate likelihood which leads  asymptotically to unbiased estimating equations obtained by the derivatives of the composite log-likelihoods. Estimation of the model parameters
can be approached by solving the marginal estimating equations
 or equivalently by
maximizing the sum of composite (univariate) likelihoods. 

By using composite likelihood the authors are assuming between-study independence in sensitivities and specificities and thus the joint likelihood in (\ref{mixed-cop-likelihood}) reduces to:   
\begin{equation}\label{ind-mixed-cop-likelihood}
L(\pi_1,\pi_2,\de_1,\de_2)=\prod_{i=1}^N\int_{0}^{1}\int_{0}^{1}
\prod_{j=1}^2g\Bigl(y_{ij};n_{ij},l^{-1}\bigl(F^{-1}(u_j;l(\pi_j),\de_j)\bigr)\Bigr)du_1du_2
=L_1(\pi_1,\de_1)L(\pi_2,\de_2),
\end{equation}
where $L_j(\pi_j,\de_j)=\prod_{i=1}^N\int_{0}^{1}g\Bigl(y_{ij};n_{ij},l^{-1}\bigl(F^{-1}(u_j;l(\pi_j),\de_j)\bigr)\Bigr)du_j$,
since under the independence assumption the copula density $c(\cdot)$ is equal to 1. Note that the joint likelihood reduces to the product of two univariate likelihoods and the evaluation of univariate integrals, thus the computational effort (if any) is subsided. Essentially, for  
beta margins  the univariate likelihoods $L_j,\,j=1,2$ result in a closed form since 
$$
\int_{0}^{1}g\Bigl(y;n,F^{-1}(u;\pi,\g_1\bigr)\Bigr)du=
\int_{0}^{1}g(y;n,x)\,dF(x;\pi,\g)=h(y;n,\pi,\g),
$$
where 
$$h(y;n,\pi,\g)=\binom{n}{y}\frac{B\Bigl(y+\pi/\g-\pi,n-y+(1 - \pi)(1 - \g)/\g\Bigr)}{B\Bigl(\pi/\g-\pi,(1 - \pi)(1 - \g)/\g\Bigr)},\, y=0,1,\ldots,n,\, 0<\pi,\g< 1,$$
is the pmf of a Beta-Binomial($n,\pi,\g$) distribution.

Composite likelihood estimates can be obtained  by maximizing the logarithm of the joint likelihood in (\ref{ind-mixed-cop-likelihood}) over the univariate parameters.
The efficiency of the composite likelihood estimates  has been studied and shown in a series of a papers \citep{varin08,Varin-etal2011}.   However,  CL ignores the dependence at the estimation of the univariate marginal parameters, thus it is expected to be worse as the dependence increases. 

\subsubsection{Asymptotic covariance matrix--Inverse Godambe}
Let $\a=(\pi,\de)$. The asymptotic covariance matrix for the CL estimator $(\a_1,\a_2)$, also known as the inverse Godambe
information matrix \citep{godambe91}, is

\begin{equation}\label{godambe}
\V=\begin{pmatrix}
\I_{11}^{-1} & \I_{11}^{-1}\I_{12}I_{22}^{-1}\\
(\I_{11}^{-1}\I_{12}I_{22}^{-1})^\top& \I_{22}^{-1}
\end{pmatrix},
\end{equation}
where $\I_{jj}=E\bigl[-\partial^2 \log L_j(\a_j)/\partial \a_j^2\bigr],\,j=1,2$ and $\I_{12}=E\Bigl[\frac{\partial\log L_1(\a_1)}{\partial \a_1}\frac{\partial\log L_2(\a_2)}{\partial \a_2}^\top\Bigr].$ For more information, including the observed inverse Godambe information matrix, we refer  the reader to \cite{Chen-etal-smmr-2014,Chen-etal-sim-2015b}.

\section{\label{sec-sim}Small- and moderate-sample efficiency--misspecification of the univariate distribution of the random effect}

In this section  an extensive  simulation study with two  different scenarios is conducted (a) to assess the performance of the CL and ML methods, and (b) to investigate in detail the effect of the misspecification of the parametric margin  of the random effects distribution.  The CL method assumes the independence copula and its focus is on marginal parameters and apparently not the choice of the copula. Hence in the simulations  we only investigate  the effect of the misspecification of the parametric margin of the random effects distribution.  We  refer the interested reader to \cite{Nikoloulopoulos2015b}  for a  study on the  misspecification of the parametric family of copulas of the random effects distribution. 

 We  use the simulation  process in \cite{Nikoloulopoulos2015b} and set the univariate parameters and disease prevalence to  mimic the telomerase  data in Section \ref{sec-app}.   The details are given below:
\begin{enumerate}
\itemsep=5pt
\item Simulate the study size $n$ from a shifted gamma distribution, i.e., $n\sim \mbox{sGamma}(\a=1.2,\b=0.01,\mbox{lag}=30)$ and round off to the nearest integer.  
 \item Simulate $(u_1,u_2)$ from a parametric family of copulas $C(;\tau)$;  $\tau$ is converted 
to the  dependence parameter $\th$ via the relations  in Table \ref{2fam}.  

\item Convert to beta  or normal realizations via $x_j=l^{-1}\Bigl(F_j^{-1}\bigl(u_j,l(\pi_j),\de_j\bigr)\Bigr)$ for $j=1,2$. 
\item Draw the number of diseased $n_{1}$ from a $B(n,0.534)$ distribution.
\item Set  $n_2=n-n_1$ and generate $y_j$ from a $B(n_j,x_j)$ for $j=1,2$. 
\end{enumerate}

In the first scenario  the simulated data are generated from the BVN copula mixed model with normal margins, logit link (the resulting model  is the same with the  GLMM) and true marginal parameters $(\pi_1,\pi_2,\s_1,\s_2)=(0.79,0.91,0.43,1.83)$, while in the second scenario  the simulated data are generated from the BVN copula mixed model with beta margins and true marginal parameters $(\pi_1,\pi_2,\g_1,\g_2)=(0.76,0.81,0.03,0.28)$. The number of studies is set to $N=10$ and $N=20$ to represent a relatively small and moderate meta-analysis, and the Kendall's $\tau$ association between study-specific sensitivity and specificity is set to $\tau=-0.5$ and $\tau=-0.8$  to represent moderate and strong negative dependence.

\setlength{\tabcolsep}{10pt}
\begin{table}[!h]
  \centering
  \caption{\label{non-conv}Times of non-convergence out of $10^4$ simulations for the CL and ML methods under different marginal choices in both simulated scenarios.}
    \begin{tabular}{ccccccccc}
    \hline
True margin  &       & $N$ & $\tau$ &       &ML-normal & ML-beta  & CL-normal & CL-beta \\
 \cmidrule{1-1}  \cmidrule{3-4} \cmidrule{6-9}
    normal &       & 10    & -0.5  &       & 45    & 17    & 995   & 0 \\
          &       &       & -0.8  &       & 14    & 7     & 1025  & 0 \\
          &       & 20    & -0.5  &       & 3     & 0     & 1118  & 0 \\
          &       &       & -0.8  &       & 1     & 0     & 1136  & 0 \\
          &       &       &       &       &       &       &       &  \\
    beta  &       & 10    & -0.5  &       & 50    & 27    & 1384  & 0 \\
          &       &       & -0.8  &       & 24    & 8     & 1495  & 0 \\
          &       & 20    & -0.5  &       & 0     & 0     & 1575  & 0 \\
          &       &       & -0.8  &       & 0     & 0     & 1576  & 0 \\
   
    \hline
    \end{tabular}
\end{table}

\setlength{\tabcolsep}{3pt}

\begin{sidewaystable}
  \sf\centering
  \caption{\label{norm-sim}
Biases,  root mean square errors (RMSE) and standard deviations (SD), along with the square root of the average theoretical variances ($\sqrt{\bar V}$) for ML and CL estimates under different margins.   
}
    \begin{tabular}{ccccccccccccccccccccccc}
   \hline
          &       &       &       & \multicolumn{4}{c}{ML-normal} &       & \multicolumn{4}{c}{ML-beta}   &       & \multicolumn{4}{c}{CL-normal} &       & \multicolumn{4}{c}{CL-beta} \\
     \cmidrule{5-8} \cmidrule{10-13} \cmidrule{15-18} \cmidrule{20-23}   
          & $N$ & $\tau$ &       & Bias  & SD    & $\sqrt{\bar V}$ & RMSE  &       & Bias  & SD    & $\sqrt{\bar V}$ & RMSE  &       & Bias  & SD    & $\sqrt{\bar V}$ & RMSE  &       & Bias  & SD    & $\sqrt{\bar V}$ & RMSE \\\hline

    $\pi_1$ & 10    & -0.5  &       & 0.08  & 2.95  & 2.62  & 2.95  &       & -0.68 & 2.91  & 2.63  & 2.99  &       & 0.02  & 2.93  & 2.80  & 2.93  &       & -0.67 & 2.91  & 2.72  & 2.98 \\
          &       & -0.8  &       & 0.01  & 2.96  & 2.52  & 2.96  &       & -0.82 & 2.93  & 2.65  & 3.04  &       & -0.03 & 2.95  & 2.80  & 2.95  &       & -0.72 & 2.91  & 2.75  & 3.00 \\
          & 20    & -0.5  &       & 0.01  & 2.07  & 1.91  & 2.07  &       & -0.74 & 2.03  & 1.93  & 2.16  &       & -0.01 & 2.07  & 2.00  & 2.07  &       & -0.77 & 2.05  & 1.98  & 2.19 \\
          &       & -0.8  &       & 0.11  & 2.09  & 1.81  & 2.10  &       & -0.74 & 2.07  & 1.89  & 2.19  &       & 0.08  & 2.08  & 1.99  & 2.08  &       & -0.67 & 2.06  & 1.98  & 2.17 \\
          &       &       &       &       &       &       &       &       &       &       &       &       &       &       &       &       &       &       &       &       &       &  \\
    $\pi_2$ & 10    & -0.5  &       & -1.76 & 5.87  & 4.38  & 6.13  &       & -9.12 & 7.04  & 4.72  & 11.52 &       & -1.68 & 5.57  & 5.13  & 5.82  &       & -9.29 & 6.99  & 5.76  & 11.63 \\
          &       & -0.8  &       & -1.60 & 5.87  & 4.13  & 6.08  &       & -8.69 & 7.06  & 4.54  & 11.20 &       & -1.61 & 5.63  & 5.17  & 5.86  &       & -9.18 & 7.05  & 5.76  & 11.58 \\
          & 20    & -0.5  &       & -0.93 & 3.90  & 3.18  & 4.01  &       & -9.40 & 5.04  & 3.72  & 10.67 &       & -0.92 & 3.74  & 3.63  & 3.85  &       & -9.43 & 5.01  & 4.30  & 10.68 \\
          &       & -0.8  &       & -1.01 & 3.97  & 3.10  & 4.09  &       & -9.16 & 5.01  & 3.50  & 10.44 &       & -0.96 & 3.75  & 3.64  & 3.87  &       & -9.48 & 4.99  & 4.31  & 10.71 \\
          &       &       &       &       &       &       &       &       &       &       &       &       &       &       &       &       &       &       &       &       &       &  \\
    $\s_1$ & 10    & -0.5  &       & -6.16 & 16.46 & 15.55 & 17.57 &       &       & 1.94  & 1.86  &       &       & -8.39 & 18.59 & 56.84 & 20.40 &       &       & 1.97  & 1.99  &  \\
          &       & -0.8  &       & -4.66 & 15.36 & 14.38 & 16.05 &       &       & 1.88  & 1.83  &       &       & -8.71 & 18.52 & 61.11 & 20.47 &       &       & 1.94  & 2.03  &  \\
          & 20    & -0.5  &       & -3.42 & 11.75 & 10.82 & 12.24 &       &       & 1.48  & 1.38  &       &       & -3.99 & 12.72 & 19.78 & 13.33 &       &       & 1.48  & 1.43  &  \\
          &       & -0.8  &       & -2.97 & 10.49 & 9.84  & 10.90 &       &       & 1.40  & 1.31  &       &       & -4.35 & 12.67 & 21.13 & 13.39 &       &       & 1.45  & 1.43  &  \\
          &       &       &       &       &       &       &       &       &       &       &       &       &       &       &       &       &       &       &       &       &       &  \\
    $\s_2$ & 10    & -0.5  &       & -20.52 & 46.38 & 42.88 & 50.72 &       &       & 10.54 & 7.42  &       &       & -20.85 & 45.98 & 14.51 & 50.49 &       &       & 10.20 & 8.68  &  \\
          &       & -0.8  &       & -21.28 & 43.66 & 40.11 & 48.57 &       &       & 10.13 & 6.84  &       &       & -20.76 & 45.96 & 14.49 & 50.43 &       &       & 10.17 & 8.70  &  \\
          & 20    & -0.5  &       & -12.33 & 32.85 & 31.25 & 35.09 &       &       & 7.81  & 5.82  &       &       & -12.12 & 32.71 & 10.02 & 34.89 &       &       & 7.61  & 6.62  &  \\
          &       & -0.8  &       & -13.80 & 31.22 & 28.48 & 34.14 &       &       & 7.65  & 5.16  &       &       & -11.94 & 32.57 & 10.01 & 34.69 &       &       & 7.56  & 6.64  &  \\
          &       &       &       &       &       &       &       &       &       &       &       &       &       &       &       &       &       &       &       &       &       &  \\
    dOR   & 10    & -0.5  &       & 14.46 & 41.89 & 35.65 & 44.32 &       & 51.62 & 54.24 & 36.04 & 74.88 &       & 13.46 & 39.07 & 33.66 & 41.32 &       & 52.46 & 53.57 & 43.79 & 74.98 \\
          &       & -0.8  &       & 14.65 & 45.93 & 35.06 & 48.21 &       & 49.71 & 57.06 & 36.70 & 75.68 &       & 13.93 & 42.23 & 36.62 & 44.47 &       & 52.69 & 56.20 & 45.95 & 77.04 \\
          & 20    & -0.5  &       & 6.87  & 23.95 & 24.03 & 24.92 &       & 48.18 & 35.32 & 26.27 & 59.74 &       & 6.61  & 22.75 & 21.32 & 23.69 &       & 48.11 & 34.95 & 30.53 & 59.46 \\
          &       & -0.8  &       & 8.10  & 25.89 & 24.52 & 27.13 &       & 47.48 & 36.61 & 26.27 & 59.95 &       & 7.47  & 24.07 & 22.60 & 25.20 &       & 49.41 & 36.59 & 32.12 & 61.48 \\
   
\hline
\end{tabular}

\begin{footnotesize}
\begin{flushleft}
\noindent 
The data are generated from the BVN copula mixed model with normal margins (GLMM) with true marginal parameters $(\pi_1,\pi_2,\s_1,\s_2)=(0.79,0.91,0.43,1.83)$ 
for different number of studies $N$ and different values of Kendall's $\tau$ association between study-specific sensitivity and specificity.  All entries are multiplied by 100. 
\end{flushleft}
  \end{footnotesize}  

\end{sidewaystable}

\begin{sidewaystable}
  \sf\centering
  \caption{\label{beta-sim} 
Biases,  root mean square errors (RMSE) and standard deviations (SD), along with the square root of the average theoretical variances ($\sqrt{\bar V}$)  for ML and CL estimates under different margins. }

    \begin{tabular}{ccccccccccccccccccccccc}
   \hline
          &       &       &       & \multicolumn{4}{c}{ML-normal} &       & \multicolumn{4}{c}{ML-beta}   &       & \multicolumn{4}{c}{CL-normal} &       & \multicolumn{4}{c}{CL-beta} \\
     \cmidrule{5-8} \cmidrule{10-13} \cmidrule{15-18} \cmidrule{20-23}   
          & $N$ & $\tau$ &       & Bias  & SD    & $\sqrt{\bar V}$ & RMSE  &       & Bias  & SD    & $\sqrt{\bar V}$ & RMSE  &       & Bias  & SD    & $\sqrt{\bar V}$ & RMSE  &       & Bias  & SD    & $\sqrt{\bar V}$ & RMSE \\\hline

    $\pi_1$ & 10    & -0.5  &       & 0.91  & 2.92  & 2.59  & 3.05  &       & 0.15  & 2.86  & 2.62  & 2.86  &       & 0.87  & 2.91  & 2.77  & 3.04  &       & 0.20  & 2.84  & 2.72  & 2.85 \\
          &       & -0.8  &       & 1.05  & 2.89  & 2.46  & 3.07  &       & 0.20  & 2.82  & 2.59  & 2.83  &       & 1.02  & 2.91  & 2.73  & 3.08  &       & 0.38  & 2.84  & 2.70  & 2.87 \\
          & 20    & -0.5  &       & 0.94  & 2.08  & 1.90  & 2.29  &       & 0.18  & 2.03  & 1.92  & 2.04  &       & 0.93  & 2.08  & 1.99  & 2.28  &       & 0.21  & 2.03  & 1.98  & 2.04 \\
          &       & -0.8  &       & 0.97  & 2.04  & 1.78  & 2.26  &       & 0.08  & 2.00  & 1.88  & 2.00  &       & 0.96  & 2.06  & 1.98  & 2.27  &       & 0.25  & 2.00  & 1.98  & 2.01 \\
          &       &       &       &       &       &       &       &       &       &       &       &       &       &       &       &       &       &       &       &       &       &  \\
    $\pi_2$ & 10    & -0.5  &       & 7.37  & 6.27  & 4.75  & 9.68  &       & 0.12  & 6.72  & 5.18  & 6.72  &       & 7.47  & 6.10  & 5.58  & 9.65  &       & -0.08 & 6.70  & 6.00  & 6.70 \\
          &       & -0.8  &       & 7.12  & 6.33  & 4.64  & 9.53  &       & 0.34  & 6.64  & 4.89  & 6.65  &       & 7.28  & 6.17  & 5.66  & 9.54  &       & -0.25 & 6.68  & 6.02  & 6.69 \\
          & 20    & -0.5  &       & 8.25  & 4.20  & 3.50  & 9.25  &       & -0.11 & 4.69  & 4.08  & 4.69  &       & 8.29  & 4.06  & 3.99  & 9.23  &       & -0.20 & 4.65  & 4.47  & 4.65 \\
          &       & -0.8  &       & 8.04  & 4.29  & 3.41  & 9.11  &       & 0.33  & 4.72  & 3.76  & 4.74  &       & 8.23  & 4.14  & 4.00  & 9.22  &       & -0.19 & 4.71  & 4.46  & 4.71 \\
          &       &       &       &       &       &       &       &       &       &       &       &       &       &       &       &       &       &       &       &       &       &  \\
    $\g_1$ & 10    & -0.5  &       &       & 14.84 & 13.66 &       &       & -0.44 & 1.80  & 1.71  & 1.85  &       &       & 16.59 & 67.17 &       &       & -0.49 & 1.87  & 1.84  & 1.94 \\
          &       & -0.8  &       &       & 13.73 & 12.62 &       &       & -0.45 & 1.65  & 1.63  & 1.71  &       &       & 16.74 & 38.99 &       &       & -0.63 & 1.82  & 1.83  & 1.92 \\
          & 20    & -0.5  &       &       & 10.52 & 9.71  &       &       & -0.28 & 1.32  & 1.27  & 1.35  &       &       & 11.12 & 15.20 &       &       & -0.29 & 1.35  & 1.34  & 1.38 \\
          &       & -0.8  &       &       & 9.57  & 8.83  &       &       & -0.28 & 1.25  & 1.19  & 1.28  &       &       & 11.20 & 17.56 &       &       & -0.34 & 1.35  & 1.34  & 1.39 \\
          &       &       &       &       &       &       &       &       &       &       &       &       &       &       &       &       &       &       &       &       &       &  \\
    $\g_2$ & 10    & -0.5  &       &       & 45.35 & 48.49 &       &       & -5.05 & 9.33  & 8.41  & 10.61 &       &       & 45.99 & 15.16 &       &       & -4.86 & 9.24  & 9.49  & 10.44 \\
          &       & -0.8  &       &       & 41.46 & 44.93 &       &       & -5.73 & 8.92  & 7.59  & 10.61 &       &       & 46.86 & 15.13 &       &       & -4.90 & 9.27  & 9.53  & 10.49 \\
          & 20    & -0.5  &       &       & 32.03 & 35.44 &       &       & -2.89 & 6.79  & 6.56  & 7.38  &       &       & 32.46 & 10.50 &       &       & -2.77 & 6.75  & 7.12  & 7.30 \\
          &       & -0.8  &       &       & 29.41 & 31.67 &       &       & -3.54 & 6.63  & 5.66  & 7.52  &       &       & 32.87 & 10.52 &       &       & -2.89 & 6.78  & 7.11  & 7.37 \\
          &       &       &       &       &       &       &       &       &       &       &       &       &       &       &       &       &       &       &       &       &       &  \\
    dOR   & 10    & -0.5  &       & -24.77 & 39.17 & 34.13 & 46.35 &       & 7.24  & 46.20 & 34.94 & 46.76 &       & -25.53 & 38.03 & 32.81 & 45.81 &       & 8.44  & 46.23 & 39.33 & 47.00 \\
          &       & -0.8  &       & -22.20 & 42.84 & 35.09 & 48.25 &       & 7.20  & 48.26 & 35.55 & 48.80 &       & -23.32 & 41.45 & 35.22 & 47.56 &       & 11.10 & 49.40 & 41.65 & 50.63 \\
          & 20    & -0.5  &       & -31.65 & 22.93 & 23.52 & 39.08 &       & 4.69  & 29.91 & 25.76 & 30.27 &       & -31.95 & 22.13 & 20.90 & 38.86 &       & 5.19  & 29.83 & 27.79 & 30.28 \\
          &       & -0.8  &       & -30.13 & 24.88 & 23.90 & 39.08 &       & 2.58  & 31.31 & 25.24 & 31.42 &       & -31.20 & 23.84 & 22.36 & 39.27 &       & 5.83  & 31.76 & 28.94 & 32.29 \\
    \hline
    \end{tabular}
\begin{footnotesize}
\begin{flushleft}
\noindent 
The data are generated from the BVN copula mixed model with beta margins with true marginal parameters $(\pi_1,\pi_2,\g_1,\g_2)=(0.76,0.81,0.03,0.28)$ 
for different number of studies $N$ and different values of Kendall's $\tau$ association between study-specific sensitivity and specificity.
All entries are multiplied by 100. 
\end{flushleft}
  \end{footnotesize}  
 
\end{sidewaystable}

As stated in  \cite{Chen-etal-smmr-2014,Chen-etal-sim-2015b} one advantage of the CL method is that the problem of non-convergence is avoided, so we also  report on the non-convergence of different methods in  Table \ref{non-conv}. To  summarize the simulated data, we report  the
resultant biases, root mean square errors (RMSE), and standard deviations (SD), along with average theoretical variances for the ML and CL estimates of the univariate parameters under different marginal choices based on iterations in which all four competing approaches converged in Table \ref{norm-sim} and Table \ref{beta-sim}.  
Following \cite{Chen-etal-sim-2015b} we also summarize the diagnostic odds ratio, that is dOR$={\frac{\pi_1}{(1-\pi_1)}}/{\frac{\pi_2}{(1-\pi_2)}}$. Clearly, this is a function of the univariate parameters;  its value ranges from zero to infinity, with a higher value indicating better discriminatory power.

Conclusions from the values in the table are the following:
\begin{itemize}
\itemsep=5pt
\item The CL method is nearly as efficient as the `gold standard' ML method. 

\item The meta-analytic ML and CL estimates and SDs are not robust to the margin misspecification. 

\item The ML method has negligible non-convergence issues. 

\item The CL method in \cite{Chen-etal-smmr-2014} has a non-convergence rate between 10\% to 16\%. 

\item The CL method in \cite{Chen-etal-sim-2015b} has  no non-convergence issues at all as expected since the $\log L$ has a closed form. 
\end{itemize}

The simulation results indicate that for both methods the effect of misspecifying the marginal choice can be seen  as substantial for both the univariate parameters and the parameters that are functions of them, such as the dOR. 
This is in line with \cite{Nikoloulopoulos2015b,Nikoloulopoulos2015c} for the ML method. Here we also show that the CL method is not robust to the misspecification of the margin.   This agrees with the conclusions of  \cite{Xu&Reid-2011} and \cite{Ogden2016} who argue that if the marginal distribution of the random effects is misspecified then the CL estimator no longer retains robustness.  This has not been revealed before, since  \cite{Chen-etal-smmr-2014}  \citep{Chen-etal-sim-2015b} focused solely on a beta  (normal) margin and didn't study the effect of misspecification of the marginal random effect distribution. 
The focus in the CL method is on marginal parameters and their functions (e.g., dOR).  Since these are  univariate inferences, all that matters, as regard as to the bias, is the univariate model.

\section{\label{sec-app}Tumor markers for bladder cancer}
In this section we illustrate the methods  with data of the  published meta-analyses in \cite{glas-etal-2003}; also analysed in \cite{Chen-etal-sim-2015b}. This meta-analyses deal with the most common urological cancer, that is  
bladder cancer. Several diagnostic markers are assessed including the cytology ($N=26$) which is the classical marker for detecting bladder cancer since 1945 and is not expensive compared with the reference standard (that is cystoscopy procedure), but lacks the diagnostic sensitivity. The other markers under investigation to give a better sensitivity are NMP22 ($N=14$), BTA ($N=6$), BTASTAT($N=8$), telomerase ($N=10$), and BTATRAK ($N=5$). 

For all the meta-analyses, we fit the copula mixed model for all different choices of parametric families of copulas  and margins. Sufficient choices of copulas are BVN, Frank, Clayton, and the rotated versions of the latter (Table \ref{2fam}). These families have
different strengths of tail behaviour; for more details see  \cite{Nikoloulopoulos2015b,Nikoloulopoulos2015c}.
We  use the  log-likelihood at  estimates as a rough diagnostic measure for goodness of fit between the  models and summarize the choice of the copula and margin with the largest log-likelihood,  along with the GLMM (BVN copula mixed model with normal margins) as a benchmark. We also estimate the model parameters with the CL method under the assumption of both normal (CL-norm) and beta (CL-beta) margins. In Table \ref{app-results} we report the resulting maximized ML and CL log-likelihoods, estimates, and standard errors.

\setlength{\tabcolsep}{6pt}

\begin{sidewaystable}
  \sf\centering
  \caption{\label{app-results}Estimated parameters, standard errors (SE) and log-likelihood values using the ML and CL methods  for bladder  cancer data.}
    \begin{tabular}{ccccccccccccccccccccc}
    \hline
     \multicolumn{10}{c}{NMP22}                                             & & \multicolumn{10}{c}{Telomerase} \\
     \cmidrule{1-10} \cmidrule{12-21}
    {} & \multicolumn{2}{c}{GLMM} & \multicolumn{2}{c}{CL-norm} & {} & \multicolumn{2}{c}{Cln270-beta} & \multicolumn{2}{c}{CL-beta} & {} & {} & \multicolumn{2}{c}{GLMM} & \multicolumn{2}{c}{CL-norm} &       & \multicolumn{2}{c}{BVN-beta} & \multicolumn{2}{c}{CL-beta} \\
    {} & {Est. } & {SE} & {Est. } & {SE} & {} & {Est. } & {SE} & {Est. } & {SE} & {} & {} & Est.  & SE    & Est.  & SE    &       & Est.  & SE    & Est.  & SE \\\hline
    {$\pi_1$} & {0.71} & {0.04} & {0.71} & {0.04} & {$\pi_1$} & {0.69} & {0.04} & {0.69} & {0.04} & {} & {$\pi_1$} & 0.77  & 0.03  & 0.77  & 0.03  & $\pi_1$ & 0.76  & 0.03  & 0.76  & 0.03 \\
    {$\pi_2$} & {0.79} & {0.03} & {0.79} & {0.03} & {$\pi_2$} & {0.78} & {0.03} & {0.77} & {0.03} & {} & {$\pi_2$} & 0.91  & 0.05  & 0.90  & 0.05  & $\pi_2$ & 0.81  & 0.06  & 0.81  & 0.06 \\
    {$\sigma_1$} & {0.64} & {0.19} & 0.65 &0.11 & {$\gamma_1$} & {0.08} & {0.04} & {0.08} & {0.04} & {} & {$\sigma_1$} & 0.43  & 0.13  & 0.39  & 0.08  & $\gamma_1$ & 0.03  & 0.02  & 0.03  & 0.02 \\
    {$\sigma_2$} & {0.58} & {0.15} & {0.56} & {0.13} & {$\gamma_2$} & {0.05} & {0.03} & {0.05} & {0.02} & {} & {$\sigma_2$} & 1.83  & 0.40  & 1.66 & 0.14  & $\gamma_2$ & 0.28  & 0.10  & 0.24  & 0.09 \\
    {dOR} & 0.65 & 0.20 & 0.65 & 0.19 & {dOR} & 0.65 & 0.21 & 0.66 & 0.18 & {} & {dOR} & 0.31 & 0.32 & 0.36 & 0.26 & dOR   & 0.77& 0.36  & 0.73 & 0.39 \\
    {$\tau$} & {-0.17} & {0.21} & {0} & {-} & {$\tau$} & {-0.28} & {0.20} & {0} & {-} & {} & {$\tau$} & -1    & -     & 0     & -     & $\tau$ & -1    & -     & 0     & - \\
    {$\log L$} & \multicolumn{2}{c}{-93.97} & \multicolumn{2}{c}{-98.02} & {$\log L$} & \multicolumn{2}{c}{-92.96} & \multicolumn{2}{c}{-94.05} & {} & {$\log L$} & \multicolumn{2}{c}{-50.37} & \multicolumn{2}{c}{-57.40} & $\log L$ & \multicolumn{2}{c}{-51.14} & \multicolumn{2}{c}{-55.34} \\\hline
    \multicolumn{10}{c}{BTA}                                               & & \multicolumn{10}{c}{BTATRAK} \\ \cmidrule{1-10} \cmidrule{12-21}
    {} & \multicolumn{2}{c}{GLMM} & \multicolumn{2}{c}{CL-norm} & \multicolumn{2}{c}{Cln180-norm} & &\multicolumn{2}{c}{CL-beta} & {} & {} & \multicolumn{2}{c}{GLMM} & \multicolumn{2}{c}{CL-norm} &       & \multicolumn{2}{c}{Cln270-beta} & \multicolumn{2}{c}{CL-beta} \\
    {} & {Est. } & {SE} & {Est. } & {SE}   & {Est. } & {SE} && {Est. } & {SE} & {} & {} & Est.  & SE    & Est.  & SE$^a$    &       & Est.  & SE    & Est.  & SE \\\hline
    {$\pi_1$} & {0.47} & {0.10} & {0.48} & {0.10} & {0.47} & {0.10} & {$\pi_1$} & {0.49} & {0.09} & {} & {$\pi_1$} & 0.66  & 0.03  & 0.67  & -     & $\pi_1$ & 0.66  & 0.03  & 0.67  & 0.02 \\
    {$\pi_2$} & {0.80} & {0.04} & {0.81} & {0.04} &  {0.80} & {0.04} & {$\pi_2$} &{0.79} & {0.04} & {} & {$\pi_2$} & 0.69  & 0.14  & 0.70  & -     & $\pi_2$ & 0.66  & 0.09  & 0.66  & 0.10 \\
    {$\sigma_1$} & {0.82} & {0.32} & {0.84} & {0.19} &  {0.76} & {0.31} &{$\gamma_1$} & {0.13} & {0.08} & {} & {$\sigma_1$} & 0.12  & 0.13  & 0.00  & -     & $\gamma_1$ & 0.00  & 0.01  & 0.00  & 0.01 \\
    {$\sigma_2$} & {0.53} & {0.20} & {0.55} & {0.18} &  {0.48} & {0.19} & $\gamma_2$ &{0.04} & {0.03} & {} & {$\sigma_2$} & 1.20  & 0.52  & 1.23 & -     & $\gamma_2$ & 0.18  & 0.10  & 0.19  & 0.11 \\
    {dOR} & 0.22 & 0.11 & 0.23 & {0.09} & 0.21 & 0.10 & dOR & 0.25 & 0.10& {} & {dOR} & 0.87  & 0.81  & 0.88  & -     & dOR   & 1.01  & 0.55  & 1.03  & 0.49 \\
    {$\tau$} & {0.26} & {0.36} & {0} & {-} &  {0.30} & {0.25} &{$\tau$} & {0} & {-} & {} & {$\tau$} & -0.92 & 0.49  & 0     & -     & $\tau$ & -0.91 & 0.18  & 0     & - \\
    {$\log L$} & \multicolumn{2}{c}{-35.87} & \multicolumn{2}{c}{-37.57} & \multicolumn{2}{c}{-35.28} & $\log L$ & \multicolumn{2}{c}{-36.14} & {} & {$\log L$} & \multicolumn{2}{c}{-34.17} & \multicolumn{2}{c}{-34.61} & $\log L$ & \multicolumn{2}{c}{-33.61} & \multicolumn{2}{c}{-34.13} \\\hline
 \multicolumn{10}{c}{BTASTAT}          &       & \multicolumn{10}{c}{Cytology} \\\cmidrule{1-10} \cmidrule{12-21}
    {} & \multicolumn{2}{c}{GLMM} & \multicolumn{2}{c}{CL-norm} & {} & \multicolumn{2}{c}{BVN-beta} & \multicolumn{2}{c}{CL-beta} & {} & {} & \multicolumn{2}{c}{GLMM} & \multicolumn{2}{c}{CL-norm} &       & \multicolumn{2}{c}{Cln90-beta} & \multicolumn{2}{c}{CL-beta} \\
    {} & {Est. } & {SE} & {Est. } & {SE} &  & {Est. } & {SE} & {Est. } & {SE} & {} & {} & Est.  & SE    & Est.  & SE    &       & Est.  & SE    & Est.  & SE \\\hline
    {$\pi_1$} & {0.74} & {0.04} & {0.75} & {0.04} & {$\pi_1$} & {0.74} & {0.03} & {0.74} & {0.04} & {} & {$\pi_1$} & 0.56  & 0.04  & 0.56  & 0.04  & $\pi_1$ & 0.56  & 0.03  & 0.55  & 0.04 \\
    {$\pi_2$} & {0.75} & {0.05} & {0.76} & {0.05} & {$\pi_2$} & {0.73} & {0.05} & {0.74} & {0.05} & {} & {$\pi_2$} & 0.96  & 0.01  & 0.96  & 0.01  & $\pi_2$ & 0.92  & 0.02  & 0.92  & 0.02 \\
    {$\sigma_1$} & {0.36} & {0.17} & {0.40} & {0.10} & {$\gamma_1$} & {0.03} & {0.02} & {0.03} & {0.02} & {} & {$\sigma_1$} & 0.75  & 0.13  & 0.71 & 0.07  & $\gamma_1$ & 0.11  & 0.03  & 0.10  & 0.03 \\
    {$\sigma_2$} & {0.72} & {0.21} & {0.72} & {0.14} & $\gamma_2$ & {0.08} & {0.04} & {0.08} & {0.04} & {} & {$\sigma_2$} & 1.46  & 0.23  & 1.51 & 0.10  & $\gamma_2$ & 0.12  & 0.05  & 0.12  & 0.04 \\
    {dOR} & 0.94 & 0.38 & 0.94 & 0.37 & dOR & 1.02 & 0.41 & 1.01 & 0.34 & {} & {dOR} & 0.05 &0.03& 0.05& 0.02 & dOR   & 0.11 & 0.04  & 0.11 & 0.03 \\
    {$\tau$} & {-0.30} & {0.38} & {0} & {-} & {$\tau$} & {-0.29} & {0.38} & {0} & {-} & {} & {$\tau$} & -0.09 & 0.19  & 0     & -     & $\tau$ & -0.06 & 0.11  & 0     & - \\
    {$\log L$} & \multicolumn{2}{c}{-51.70} & \multicolumn{2}{c}{-54.55} & {$\log L$}& \multicolumn{2}{c}{-51.52} & \multicolumn{2}{c}{-51.84} & {} & {$\log L$} & \multicolumn{2}{c}{-153.28} & \multicolumn{2}{c}{-158.39} & $\log L$ & \multicolumn{2}{c}{-152.08} & \multicolumn{2}{c}{-152.18} \\
    \hline
    \end{tabular}%

\begin{flushleft}
\begin{footnotesize}
Cln$\omega$-norm and Cln$\omega$-beta denotes a  Clayton  rotated by $\omega$ degrees copula mixed with normal and beta margins, respectively.
\medskip

$^a$The CL-norm estimate of the between study variance $\s_1^2$  was approximately zero, thus for this case   the standard errors are unreliable as the between-study variance parameter estimate is on the boundary of the parameter space.

\end{footnotesize}
\end{flushleft}
\end{sidewaystable}

\subsubsection*{NMP22}
The log-likelihoods show that  a copula mixed model with rotated by 270 degrees Clayton copula and beta margins provides the best fit and the estimates of sensitivity $\pi_1$ and specificity $\pi_2$ are smaller under this assumption. The CL method performs well since the  estimated $\tau$ is weak and not significantly different from zero.

\subsubsection*{BTA}
The log-likelihoods show that a copula mixed model with rotated by 180 degrees Clayton copula and normal margins provides the best fit. 
\cite{Chen-etal-sim-2015b} previously restricted to beta margins thus the sensitivity $\pi_1$ and dOR were overestimated (CL-beta). 

\subsubsection*{BTASTAT}
The log-likelihoods show that a BVN copula mixed model with   beta margins provides the best fit and the estimates of specificity $\pi_2$ and dOR are smaller and larger, respectively, under this assumption. 

\subsubsection*{Telomerase}
 \cite{Nikoloulopoulos2015b} has previously analysed these data to illustrate the copula mixed model when there 
 exists negative perfect dependence, and thus there is  only one copula: the countermonotonic copula. This is a limiting case for all the parametric families of copulas, when the dependence parameter is fixed to the left boundary of its parameter space.
Both models agree on the estimated sensitivity $\hat\pi_1$ but the estimate of specificity $\hat\pi_2$ is larger under the standard GLMM.   The log-likelihood  is $-50.37$ for normal margins and $-51.14$ for beta margins, and thus a normal margin seems to be a better fit for the data. In this example the CL method overestimates the dOR,   
since it ignores the perfect negative dependence at the estimation of the parameters.

\subsubsection*{BTATRAK}
The log-likelihoods show that a copula mixed model with rotated by 270 degrees Clayton copula and beta margins provides the best fit. Note that the CL-norm estimate of the between study variance $\s_1^2$  was approximately zero, thus for this case   the standard errors are unreliable as the between-study variance parameter estimate is on the boundary of the parameter space.

\subsubsection*{Cytology}
 The log-likelihoods show that a copula mixed model with rotated by 90 degrees Clayton copula and beta margins provides the best fit. All models agree on the estimated sensitivity $\hat\pi_1$, but the estimated  specificity $\pi_2$ and dOR  are smaller when beta margins are assumed. The CL method performs well on the estimation of the univariate parameters and their functions since the  estimated $\tau$ is weak and not significantly different from zero.

\section{\label{sec-disc}Discussion}
In this paper we have challenged claims made in \cite{Chen-etal-smmr-2014,Chen-etal-sim-2015b}  about the advantages of using a composite likelihood  in meta-analysis of diagnostic test accuracy studies, in terms of convergence and robustness to model misspecification. The usual reason for using a composite likelihood does not apply here, because the full likelihood is straightforward to compute. We have demonstrated that the copula mixed model  in  
\cite{Nikoloulopoulos2015b} 
does not suffer for computational problems or convergence issues.   \cite{Nikoloulopoulos2015b} proposed a numerically stable ML estimation technique based on Gauss-Legendre quadrature; the crucial step is to convert from independent to dependent quadrature points. Furthermore, 
it has been shown the secondary motivation of robustness of the CL method  is not retained in this context if the marginal distributions are misspecified. 
Hence it is a digression to use the CL methods in \cite{Chen-etal-smmr-2014,Chen-etal-sim-2015b}  for estimation in meta-analysis of diagnostic test accuracy studies as apparently there is neither  computationally difficulty in the calculation of the bivariate log-likelihood nor robustness in the misspecification of the marginal distribution of the random effects. 
These conclusions hold to any context where clinical trials or observational studies report more than a single binary outcome.

Furtermore, in \cite{Chen-etal-smmr-2014,Chen-etal-sim-2015b}  the main inference is univariate such as the 
overall sensitivity or specificity or their functions as a single measure of diagnostic accuracy, e.g., the diagnostic odds ratio (dOR).  The dOR  for many cases is not useful since cannot distinguish the ability to detect individuals with disease from the ability to identify healthy individuals \citep{Chen-etal-smmr-2014}. 
Whenever  the balance between false negative and false positive rates is of immediate importance, both the prevalence and the conditional error rates of the test have to be taken into consideration to make a balanced decision; hence   the dOR is less useful, as it does not distinguish between the two types of diagnostic mistake \citep{Glas2003}.

\begin{figure}[!h]
\begin{center}
\begin{tabular}{|cc|}

\hline NMP22& Telomerase \\\hline

\includegraphics[width=0.4\textwidth]{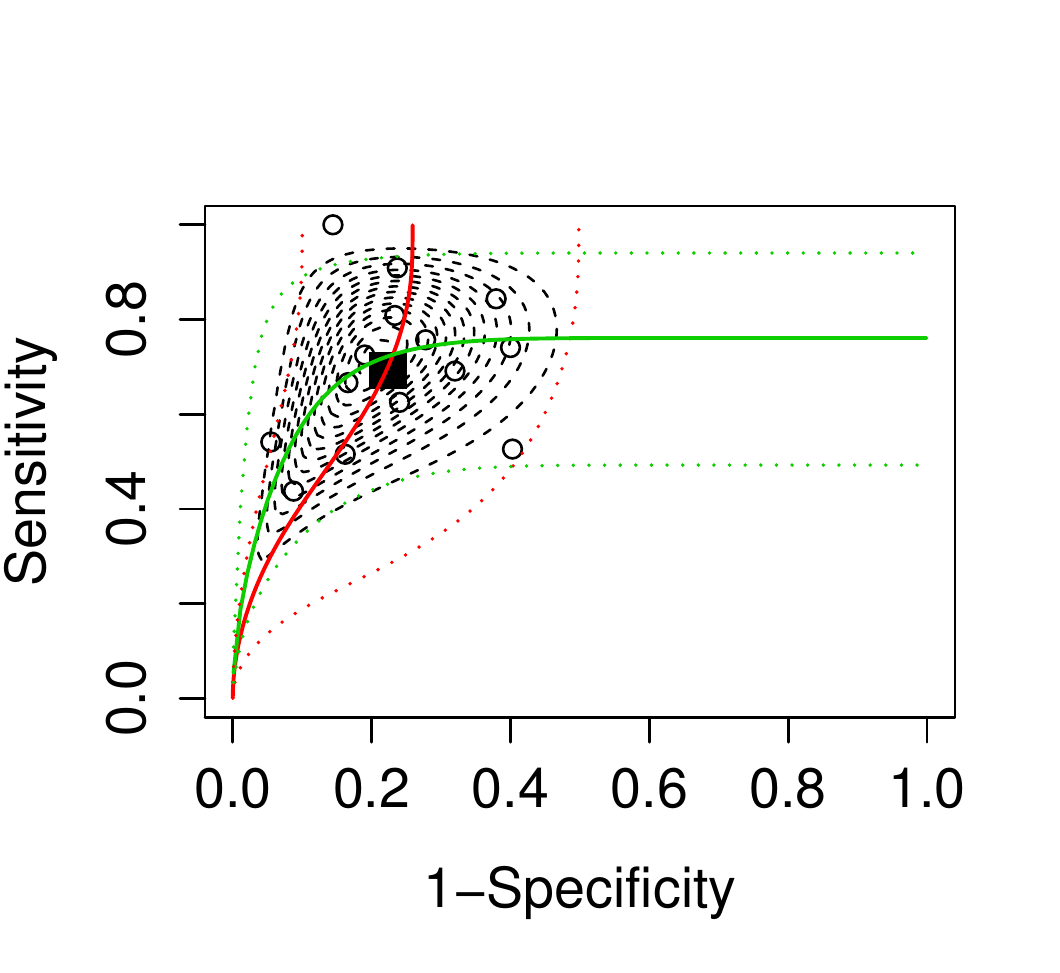}
&

\includegraphics[width=0.4\textwidth]{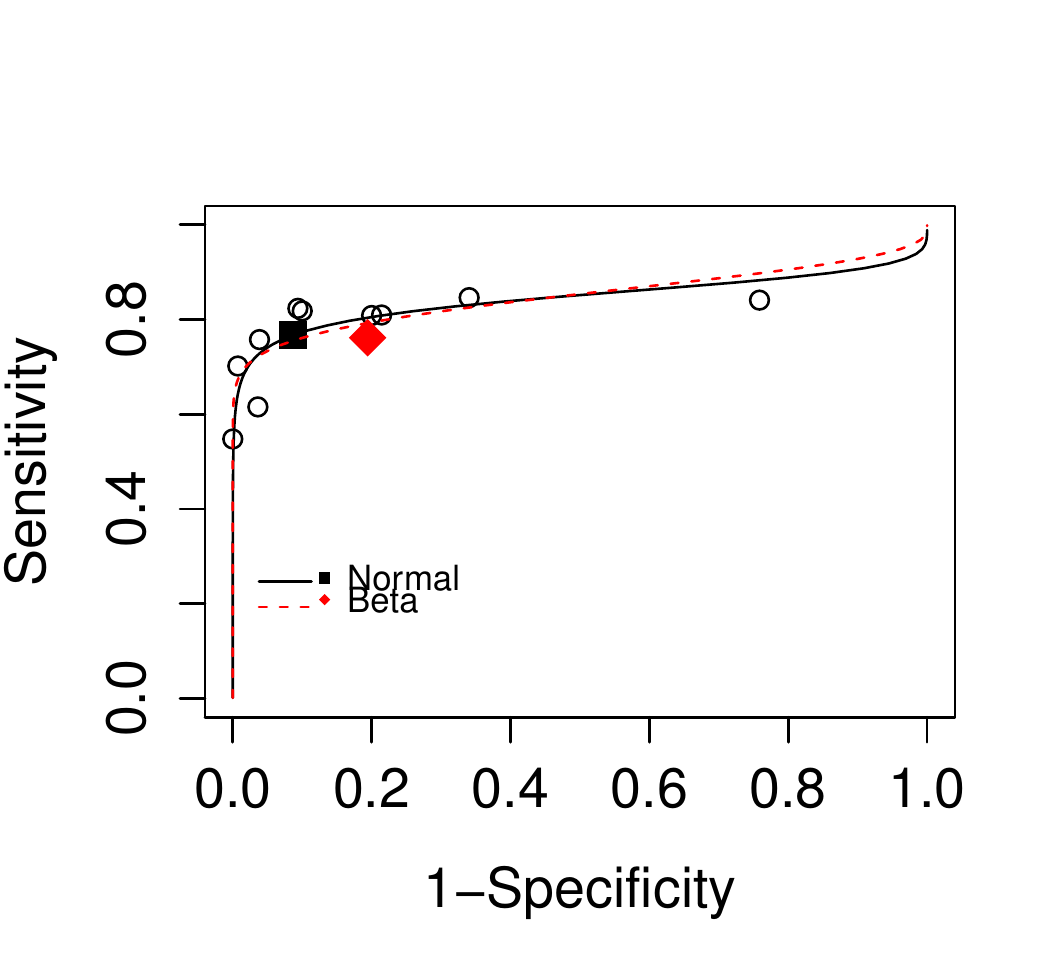}\\\hline
BTA& BTATRAK\\\hline
\includegraphics[width=0.4\textwidth]{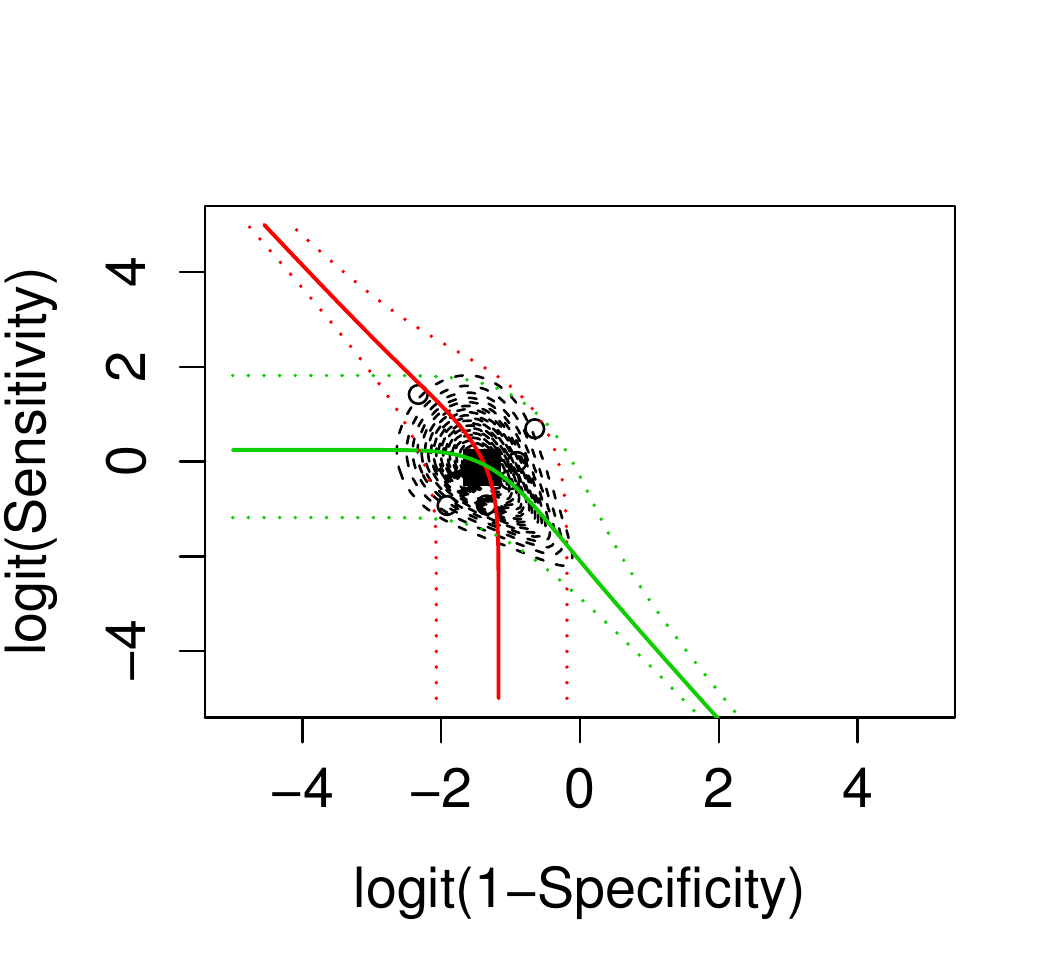}
&
\includegraphics[width=0.4\textwidth]{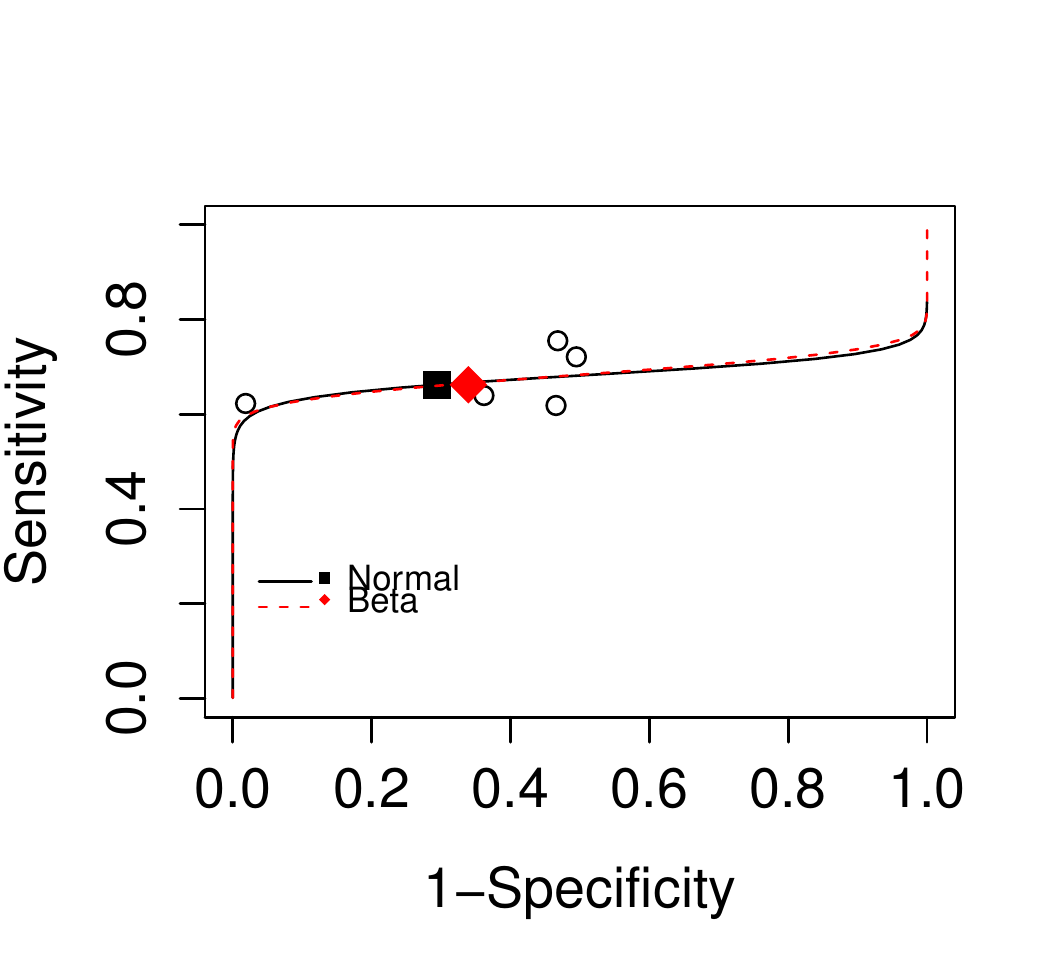}\\\hline
BTASTAT& Cytology\\\hline
\includegraphics[width=0.4\textwidth]{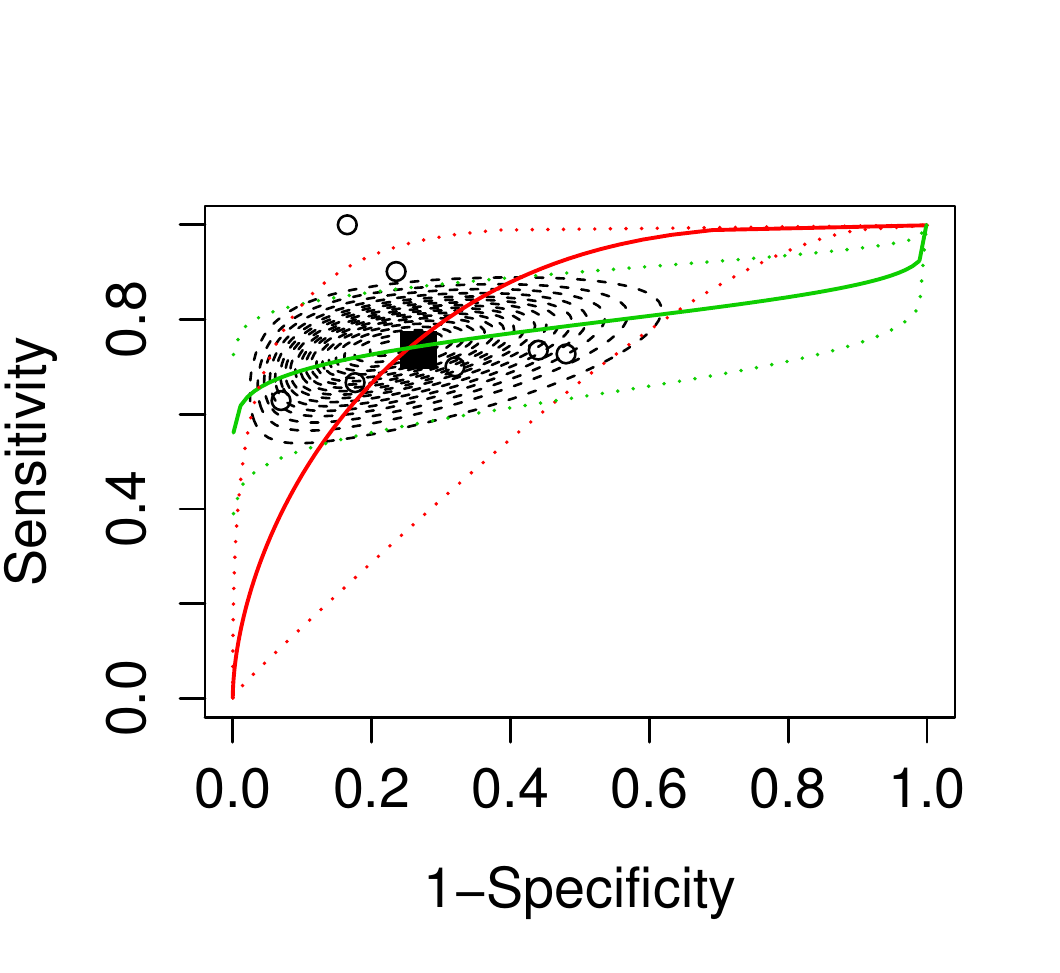}
&
\includegraphics[width=0.4\textwidth]{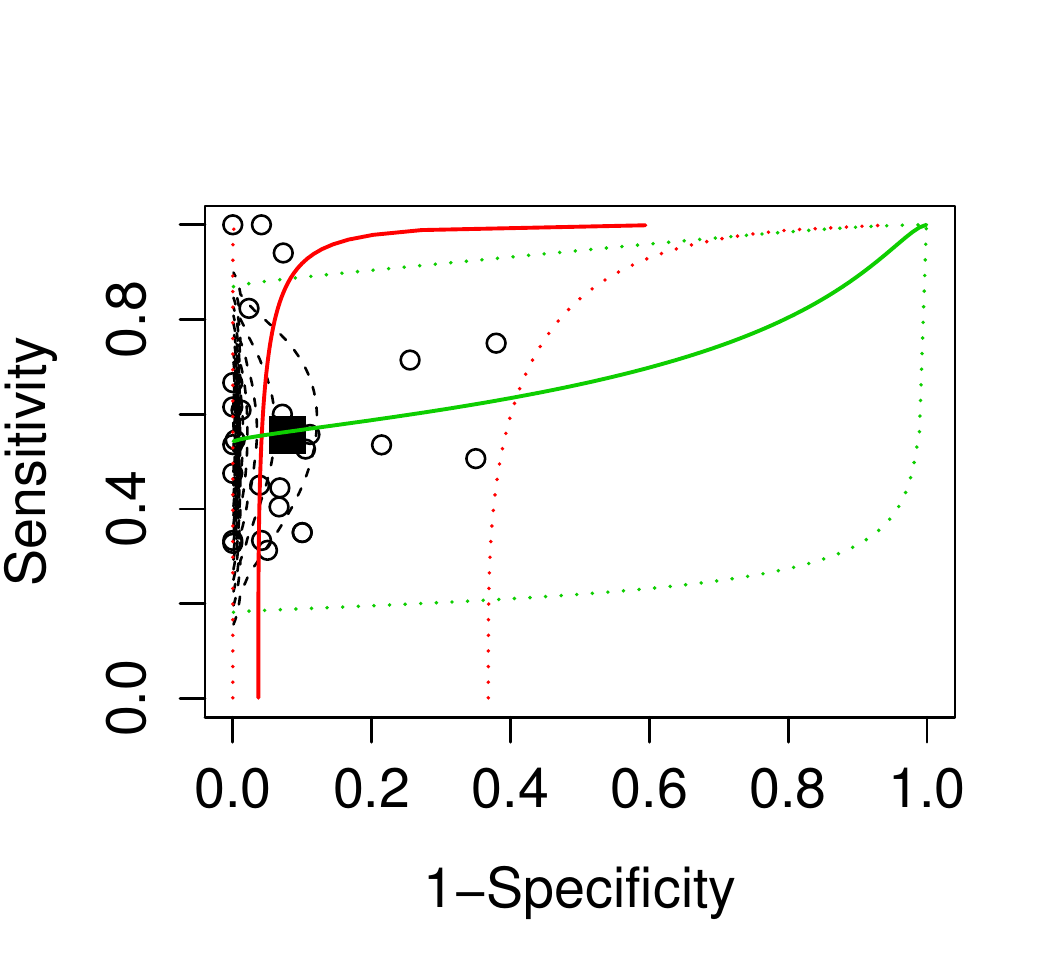}\\
\hline
\end{tabular}
\caption{\label{SROCs}Contour plots (predictive region)  and quantile  regression curves  from the best fitted copula mixed model for the  bladder  cancer data. Red and green lines represent the quantile  regression curves $x_1:=\widetilde{x}_1(x_2,q)$ and $x_2:=\widetilde{x}_2(x_1,q)$, respectively; for $q=0.5$ solid lines and for $q\in\{0.01,0.99\}$ dotted lines (confidence region). In case of BTATRAK and telomerase the predictive and confidence  region are meaningless since the  Kendall's $\tau$ association is close to $-1$.  In this case all the quantile regression curves almost coincide, and hence, we  depict only the median regression curve for each model. In case of BTA the axes are in  logit scale since  we also plot  the estimated contour plot of the random effects distribution as predictive region; this has been estimated for the logit pair of (Sensitivity, Specificity).}
\end{center}
\end{figure}

In fact, if the interest is only to overall sensitivity,  and specificity, then the overall test accuracy across studies will not be clearly defined.  Different studies use different thresholds for a positive test result, thus  the overall sensitivity and specificity  do not make sense. 
Instead, some form of the summary receiver operating characteristic (SROC) curve makes much more sense and will help decision makers to assess the actual diagnostic accuracy of a diagnostic test. 
In an era of evidence-based medicine, decision makers need high-quality procedures such as the SROC curves to support decisions about whether or not to use a diagnostic test in a specific clinical situation and, if so, which test.  

An SROC curve is deduced for the copula mixed model  in \cite{Nikoloulopoulos2015b} through a median regression  curve of $X_1$ on $X_2$. 
For the copula mixed model, the model parameters (including dependence parameters), the choice of the copula, and the choice of the margin affect the shape of the SROC curve \citep{Nikoloulopoulos2015b}.  
However, there is no priori reason to regress $X_1$ on $X_2$ instead of the other way around, so \cite{Nikoloulopoulos2015b} also provides  a median regression  curve of $X_2$ on $X_1$. 
Apparently, while there is a unique definition of the ROC curve within a study with fixed accuracy, there is no unique definition of SROC curve across multiple studies with different accuracies \citep{Rucker&Schumacher2010}. As \cite{Arends-etal-2008} have pointed out, none of the SROC curves proposed in the literature can be interpreted as an average ROC. 
 \cite{Rucker-schumacher-2009} stated that instead of summarizing data using an SROC, it might be preferable  to give confidence regions.  Hence, in addition to using just median regression curves, \cite{Nikoloulopoulos2015b} proposed quantile regression curves with a focus on high ($q$ = 0.99) and low quantiles ($q$ = 0.01), which are strongly associated with the upper and lower tail dependence imposed from each parametric family of copulas.  These can been seen as confidence regions of the median regression SROC curve.  Among the parametric families of copulas in Table \ref{2fam} the tail dependence varies, and is a property to consider when choosing amongst different families of copulas as affects the shape of SROC curves \citep{Nikoloulopoulos2015b}.
Finally, \cite{Nikoloulopoulos2015b}
to reserve the nature of a bivariate response instead of a univariate response along with a covariate,  proposed to plot the estimated contour of the random effects distribution. The contour plot can be seen as the predictive region of the estimated pair of sensitivity and specificity. The prediction region of the copula mixed model  does not depend on the assumption of bivariate normality of the random effects and has non-elliptical shape.

Figure \ref{SROCs}  demonstrates these curves and summary operating points (a pair of average sensitivity and specificity) with a confidence  and a predictive region from the best fitted  copula mixed model for all the meta-analyses in Section \ref{sec-app}. 
Both  CL methods in \cite{Chen-etal-smmr-2014,Chen-etal-sim-2015b} cannot be used to produce the SROC curves, since the  dependence parameters  affect the shape of the SROC curve and these are set to independence by definition.   Note in passing that the CL method in \cite{Chen-etal-smmr-2014} can provide a confidence region but this is restricted to the elliptical shape.

Nevertheless, the additional feature of having to estimate the association among the random effects in ML estimation has been found to require larger sample sizes than in CL estimation where this parameter is set to independence. The application example includes cases with an adequate number of individual studies. For meta-analyses with fewer  studies the CL methods in \cite{Chen-etal-smmr-2014,Chen-etal-sim-2015b} can  be recommended   if a bivariate copula mixed model is near non-identifiable (or has  a flat log-likelihood) and the estimation of an average operating point (summary sensitivity and specificity) is of interest instead of a SROC curve.

\section*{Software}
The R package {\tt CopulaREMADA}  \citep{Nikoloulopoulos-2015} has been used to produce  the ML estimates (along with their SE) of the parameters from the copula mixed models and  plot the SROC curves and summary operating points (a pair of average sensitivity and specificity) with a confidence and a predictive region.  The R package {\tt  xmeta} \citep{chen-etal-2016-R-package} has been used to  produce the CL estimates (along with their SE) of the parameters from both methods in \cite{Chen-etal-smmr-2014,Chen-etal-sim-2015b}. 

\section*{Acknowledgements}
The simulations presented in this paper were carried out on the High Performance Computing Cluster supported by the Research and Specialist Computing Support service at the University of East Anglia.


\end{document}